\documentclass[reprint,aps]{revtex4-2}

\usepackage{dcolumn}% Align table columns on decimal point
\usepackage{bm}% bold math

\usepackage[utf8]{inputenc}
\usepackage[T1]{fontenc}
\usepackage{mathptmx}
\usepackage{etoolbox}
\usepackage{amsmath}

\usepackage{graphicx}    
\usepackage{siunitx}
\usepackage{xcolor}
\usepackage{float} 
\usepackage{tabularx}
\usepackage{amssymb}
\usepackage{ulem}
\usepackage{multirow}

\begin{document}

\title{Molecular Rotors for In Situ Viscosity Mapping during Evaporation of Confined Fluid Mixtures}

\author{Florence Gibouin}
    \affiliation{Laboratoire du Futur, UMR5258 (CNRS - Universit\'e de Bordeaux - Solvay), 33600 Pessac, France}
\author{Dharshana Nalatamby}
    \affiliation{Laboratoire du Futur, UMR5258 (CNRS - Universit\'e de Bordeaux - Solvay), 33600 Pessac, France}
\author{Pierre Lidon}
    \email[corresponding author: ]{pierre.lidon@u-bordeaux.fr}
    \affiliation{Laboratoire du Futur, UMR5258 (CNRS - Universit\'e de Bordeaux - Solvay), 33600 Pessac, France}
\author{Yaocihuatl Medina-Gonzalez}
   \email[corresponding author: ]{yaocihuatl.medina-gonzalez@u-bordeaux.fr}
    \affiliation{Laboratoire du Futur, UMR5258 (CNRS - Universit\'e de Bordeaux - Solvay), 33600 Pessac, France}

\date{\today}

\begin{abstract}
Numerous formulation processes of materials involve a drying step, during which evaporation of a solvent from a multi-component liquid mixture, often confined in a thin film or in a droplet, lead to concentration and assembly of non volatile compounds. While the basic phenomena ruling evaporation dynamics are known, a precise modeling of practical situations is hindered by the lack of tools for local and time-resolved mapping of concentration fields in such confined systems. In this article, the use of Fluorescence Lifetime Imaging Microscopy and of Fluorescent Molecular Rotors is introduced as a versatile, in-situ and quantitative method to map viscosity and concentration fields in confined, evaporating liquids. More precisely, the cases of drying of a suspended liquid film and of a sessile droplet of mixtures of fructose and water is investigated. Measured viscosity and concentration fields allow to characterize drying dynamics, in agreement with simple modeling of the evaporation process.
\end{abstract}


\maketitle
%--------------------------------------------------------------------

\section{Introduction}
\label{sec:intro}

Material formulation generally involve phase changes: process indeed often start with liquid mixtures to benefit from their ability to flow for casting in a desired configuration, before a step of evaporation of solvent to obtain a solid deposit. Most of coatings start under the form of films of liquid suspensions spread over substrates~\cite{zhang_2023f}, as for paints~\cite{vanderkooij_2016,honzicek_2019} or lithium-ion batteries~\cite{zhang_2023e} among other examples. Evaporating sessile droplets over solid substrates are also an important geometry, that can be encountered in various applications like spray cooling~\cite{kim_2007}, chemical and biological assays~\cite{garciacordero_2017}, medical diagnosis~\cite{chen_2016d,hertaeg_2021,pal_2023a}, spreading of pesticides~\cite{yu_2009} or material self-assembly~\cite{heo_2023}. In the past decades, such configuration witnessed an increased interest in the context of inkjet printing~\cite{lohse_2022}, which has overcome the usual application of printing on paper~\cite{degans_2004,williams_2006} to become a cheap and versatile method for additive manufacturing, employed for flexible organic and inorganic electronics~\cite{sirringhaus_2000,kateri_2003,kawase_2003,singh_2010,stringer_2016,mattana_2017,beedasy_2020}, light-emitting devices~\cite{dijksman_2007}, pharmaceutical products~\cite{daly_2015}, biomaterials~\cite{saunders_2014}  or filtration membranes~\cite{wang_2023h}.

From a fundamental point of view, the dynamics of evaporation of a liquid is an old problem~\cite{thomas_1917a,thomas_1917b,jeffreys_1918,langmuir_1918}, and has been the subject of intense experimental and modeling efforts for decades. Evaporation dynamics is set by diffusion and convection of volatile species in the surrounding vapor, which drives temperature changes through evaporative cooling, and internal flows, either due to evaporation or to solutal and/or thermal Marangoni effect~\cite{cazabat_2010,erbil_2012,gurrala_2021,wang_2022j,gelderblom_2022,wilson_2023}. In presence of non-volatile solutes, pinning of the contact line can occur, leading to the so-called coffee-ring effect~\cite{routh_2013,larson_2014,zang_2019}, and progressive concentration can lead to crystallization at the edge of the drop~\cite{shahidzadeh_2015}. The situation is thus complex, involving diverse phenomena at multiple and entangled scales, and a comprehensive understanding is still elusive, even for the simplest case of a pure liquid. 

While temperature and velocity fields can be characterized using infrared thermal imaging and Particle Image Velocimetry for instance~\cite{dhavaleswarapu_2010,christy_2011,chen_2017}, methods for measuring local composition are scarce, despite the essential role of concentration field on the evaporation dynamics~\cite{diddens_2017a,diddens_2017b}. This absence of appropriate tools for mapping local concentration has hindered progress in the study of evaporation of multi-component systems, in particular in the case of molecular compounds that cannot be directly observed by optical microscopy. Relevant techniques should indeed allow for an in situ and minimally intrusive characterization of such confined systems. Some methods, like X-ray tomography~\cite{weon_2011} or Schlieren and infrared tomography~\cite{kellyzion_2013}, have been used to measure local density in sessile droplets and the surrounding vapor in the case of pure fluids but they cannot be used to determine concentration in the case of mixtures. Various other techniques, including infrared spectroscopy~\cite{innocenzi_2008}, Laser Induced Fluorescence and Raman scattering~\cite{hopkins_2005}, acoustic reflectometry and spectroscopy~\cite{chen_2016c,zhang_2021e}, Injected Gas Chromatography~\cite{kita_2018}, refractometry~\cite{ozturk_2020} and tensiometry~\cite{raj_2010} allowed to determine the evolution of concentration in time, yet without any spatial resolution. Finally, very few studies managed to measure concentration fields with both spatial and temporal resolution: they for instance used spatially resolved NMR~\cite{kresse_2019,kresse_2021} possibly coupled with Raman spectroscopy~\cite{bell_2022}, interferometry to detect changes of refractive index (mostly in the vapor surrounding the droplet)~\cite{dehaeck_2009,dehaeck_2014} or fluorophores sensitive to local solvent environment~\cite{cai_2017}. While these techniques offer interesting opportunities to characterize the evaporation of liquid film and droplets, they all suffer some restrictions either on the geometry of the system or on the studied compound and alternative measurement technologies are still of high interest.

In mixtures, drying is usually associated with an increase of the viscosity, in relation with progressive concentration of the non-volatile species. This makes viscosity an interesting parameter to monitor, as a proxy for concentration and as it moreover directly impacts the flows inside the fluid. Mechanical measurements have for instance been performed during drying of colloidal suspensions by measuring macroscopic viscosity with conventional rheometry~\cite{lehericey_2021} or the evolution of normal force during drying~\cite{bouchaudy_2019}, but in-situ and non-intrusive tools for viscosity characterization are lacking~\cite{barnes_1999b}. 

Fluorescent Molecular Rotors (FMR) are fluorophores whose fluorescence properties (intensity and lifetime) depend on local viscous dissipation: they thus offer an opportunity for local viscosity mapping. More precisely, their relaxation from an excited state can follow two competing paths, by undergoing either a conventional radiative relaxation, or a non-radiative relaxation associated with twisting and rotational motion of the molecule. In the latter case, interactions with the surrounding fluids, quantified by the so-called microviscosity, affect the relaxation rate, resulting in an enhanced fluorescence (intensity and lifetime) in more viscous environments, which impede rotational motion. Even if the relation between microviscosity and conventional viscosity is still unclear~\cite{gulnov_2016,bittermann_2021a,mirzahossein_2022}, both quantities are correlated. Consequently, provided a preliminary calibration of the response of fluorescence to viscosity, measurement of the fluorescence allows to retrieve spatially resolved viscosity maps~\cite{haidekker_2007,haidekker_2010,haidekker_2016}. The possibility to exploit fluorescence lifetime instead of intensity is particularly appealing. Indeed, despite requiring a Fluorescence Lifetime Imaging Microscope (FLIM) instead of a conventional fluorescence microscope, lifetime is only sensitive to local viscosity, and not on local excitation intensity or fluorophore concentration, which avoids tedious calibration steps and limits measurement errors~\cite{suhling_2005,vanmunster_2005,festy_2007}.

FMR have initially been introduced in the context of bioimaging and used to image semi-quantitatively viscosity changes in biofluids and inside cells~\cite{kung_1989,akers_2005,kuimova_2008,wu_2013,dent_2015,li_2016,briole_2021,bittermann_2023}. More recently, they have became an interesting tool to monitor physico-chemical transformations~\cite{raeburn_2015,schmitt_2022} and for the observation of nanoscopic contacts~\cite{suhina_2015,weber_2018,weber_2019,terwisschadekker_2023}. However, it has been shown that they can also be used as quantitative viscosity probes in bulk fluids~\cite{garg_2020}, in microfluidic flow~\cite{benninger_2005,bunton_2014,nalatamby_2023} or in confined lubricated flows~\cite{ponjavic_2015}. 

In this paper, lifetime mapping of FMR using FLIM is introduced as a tool to measure local viscosity fields in evaporating, confined fluid mixtures, allowing to retrieve also concentration fields. A similar study has been done in the context of aqueous aerosols, but it only characterized equilibrium properties of rather homogeneous droplets~\cite{hosny_2013}.

First, the response of a BODIPY-based FMR was calibrated in the mixtures under study, allowing to retrieve viscosity and fructose concentration from lifetime measurements. Then, measurements in suspended liquid films are reported: in this case, concentration field was homogeneous, and its dynamics is qualitatively well captured by a simple model, confirming the relevance of FMR and FLIM for such study. Finally, viscosity mapping during the evaporation of sessile droplets is described, showing the interest of the method to characterize more complex geometry, involving heterogeneous concentration profiles. 

This work thus illustrates the use of FMR and FLIM for monitoring the evolution of liquid mixtures in confined, submillimetric geometry. The method is very versatile as it only requires that the viscosity of the fluid evolves, and thus sets only little constraints on the type of system that can be characterized. It is also only little invasive, as only low FMR concentration (set to $\SI{e-5}{\mole\per\liter}$ in this study) is required. Finally, measurement is purely optical and could be adapted for inline measurements, in particular by coupling it to a confocal microscope.

\section{Material and methods}
\label{sec:mat_met}

\subsection{Preparation of solutions}
\label{subsec:solutions}

In this study, evaporation of solutions of fructose in water were studied. Fructose was chosen as being a cheap and safe material, easy to formulate and leading to solutions of high viscosity. Also, this mixture is well characterized in the literature, in particular due to its relevance in food science~\cite{cerdeirina_1997,darrosbarbosa_2003,fucaloro_2007}. Solutions were prepared by dissolving fructose (purity $\geq \SI{99}{\percent}$, Sigma Aldrich F0127) in distilled water at ambient temperature, under magnetic agitation. Solutions of various mass fraction in fructose were prepared for further calibration and experiments.

\begin{figure}[!]
    \includegraphics[width=0.4\columnwidth]{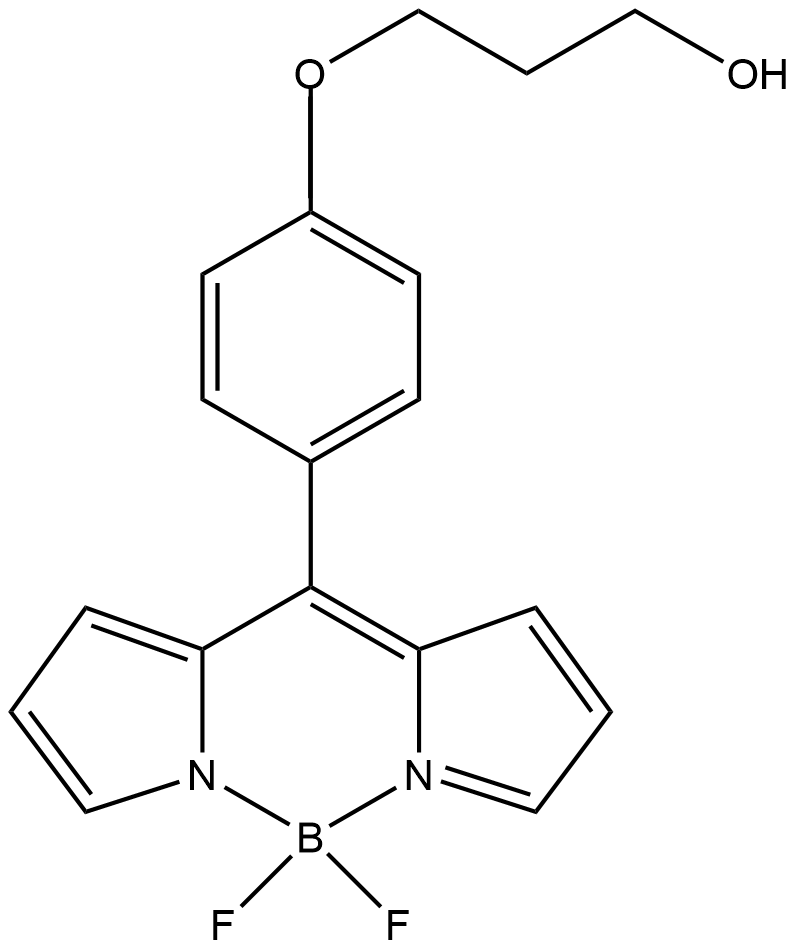}
    \caption{Structure of the BODIPY-based fluorescent molecular rotor of formula $\mathrm{C_{18}H_{17}BF_2N_2O_2}$ and molar mass $\SI{342.15}{\gram\per\mole}$, as synthesized and characterized in~\cite{nalatamby_2023}. Reprinted from Nalatamby et al.~\cite{nalatamby_2023}. Copyright 2023, Industrial \& Engineering Chemistry Research, American Chemical Society.}
    \label{fig:bodipy_structure}
\end{figure}

The FMR used in this study was a BODIPY-based compound (see structure on Fig.~\ref{fig:bodipy_structure}) which was synthesized and characterized in a previous work~\cite{nalatamby_2023}. As its solubility in aqueous solutions was limited, FMR was initially dissolved in glycerol  (fresh glycerol of purity $\geq \SI{99}{\percent}$, Sigma Aldrich G5516-1L), then diluted in the fructose/water solutions in order to reach a final mole concentration of FMR of $\SI{10}{\micro\mole\per\liter}$ in every solution. While this parameter has no impact on fluorescence lifetime measurement, it is important to keep it constant if fluorescence intensity measurements are carried out.

Consequently, the solutions under study are actually ternary mixtures of water, fructose and glycerol. Final fructose mass fractions $w_\text{f}$ of prepared solutions were between $0.5$ to $0.7$ (corresponding to molar concentration between about $\SI{0.1}{\mole\per\liter}$ and $\SI{0.2}{\mole\per\liter}$) and glycerol mass fraction  $w_\text{g}$ were between $0.04$ to $0.07$ (corresponding to molar concentration around $\SI{25}{\milli\mole\per\liter}$).

Finally, for experiments on free standing liquid films, a surfactant (Sodium Dodecyl Sulfate of purity $>\SI{99}{\percent}$, SDS, Sigma Aldrich) was added to solutions in order to stabilize the film at a mass fraction $\SI{0.1}{\percent}$ (corresponding to $\SI{0.12}{\milli\mole\per\liter}$). Considering these small amounts, presence of SDS will not be considered in further modeling.

\subsection{Fluorescence Lifetime Imaging Microscopy}
\label{subsec:FLIM}

\subsubsection{Setup}
\label{subsubsec:FLIM_setup}

\begin{figure}[!]
    \includegraphics[height=5cm]{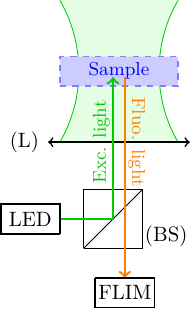}
    \caption{Schematics of the FLIM setup. The sample (in blue) is placed in the field of the objective lens (L). It is illuminated from the bottom by the excitation light of a LED source (in green, $\lambda = \SI{451}{\nano\meter}$ and fluorescent emission (in orange) is collected by a CCD camera through same objective lens, and analyzed by a computer. The different lights are separated by a dichroic beam splitter (BS).}
    \label{fig:FLIM_setup}
\end{figure}

Fluorescence lifetime $\tau$ of FMR was measured with a FLIM operating in the frequency domain (Lambert Instruments FLIM Attachment). FLIM device was mounted on an inverted microscope (Olympus IX71). According to the experiment, this microscope was equipped with different objective lenses, $1.25 \times$ (used for the study of sessile droplets) or $20 \times$ (used for suspended liquid films). A LED source of wavelength $\SI{451}{\nano\meter}$ was used close to the maximum of absorption of the FMR~\cite{nalatamby_2023}. Fluorescence emission was then collected by these objective lens and transmitted to the detector of a CCD camera, after passing through a long pass filter ($\SIrange{440}{490}{\nano\meter}$). The situation is represented on Fig.~\ref{fig:FLIM_setup}.

The CCD camera was equipped with a $504 \times \SI{512}{px^2}$ detector. Once calibrated with a calibration target, a calibration factor was obtained (respectively about $\SI{20}{\micro\meter\per px}$ and $\SI{1.25}{\micro\meter\per px}$ for $1.25 \times$ and $20 \times$ objective lens). For every sample, exposure time was adjusted as a compromise between optimizing contrast of intensity and avoiding detector saturation, while keeping acquisition rate as fast as possible. Typically, an exposure time of $\SI{900}{\milli\second}$ was used for sessile droplets and $\SI{1500}{\milli\second}$ in the case of suspended liquid films. FLIM acquisition was carried out using a modulation frequency of $f=\SI{40}{\mega\hertz}$ and $12$ acquisition phases with a $1 \times$ CCD gain. 

Lifetime measurements were calibrated using free acid fluorescein (Sigma Aldrich, 46955-1G-F) in a buffered solution  of $p\mathrm{H}=10$ with concentration $\SI{10}{\micro\mole\per\liter}$ as a reference, with tabulated lifetime $\tau_\text{ref} = \SI{4.02}{\nano\second}$~\cite{strickler_1962}. In all calibration and experiments, special care was taken to acquire the reference image in conditions as close as possible to these of the experiments~\cite{nalatamby_2023}.

Transverse spatial resolution was limited by pixel size, and thus by the magnification of the objective lens. In the direction of light propagation, lifetime is averaged on the whole depth of field of the objective lens. In the present experiments, this can be estimated to a typical depth of a few $\SI{100}{\micro\meter}$ and of a few $\SI{}{\micro\meter}$ for $1.25 \times$ and $20 \times$ objective lens. Temporal resolution was limited by exposure time and by the number of acquired phases per image: for the number of phases used here, corresponding to maximal resolution, acquisition of one full image took about $\SI{10}{\second}$.

The setup used in this study was designed to measure lifetime typically in the range $\SIrange{0.7}{5}{\nano\second}$. Resolution on lifetime measurement was principally limited by noise level, and decreased for increasing lifetime. For the experiments performed with suspended liquid films, presented in Section~\ref{subsec:results_film}, lifetime was homogeneous across the acquired images and fluctuations allowed to determine this noise level. At film formation, typical relative variations were of the order of $\SI{20}{\percent}$ for an average lifetime of $\SI{1.6}{\nano\second}$, close to the resolution limit of the apparatus. After about half hour of evolution, fluctuations were of the order of $\SI{10}{\percent}$ for an average lifetime of $\SI{2.2}{\nano\second}$. At steady state, relative fluctuations were reduced below $\SI{1}{\percent}$ for an average lifetime of $\SI{3.6}{\nano\second}$.

\subsubsection{Image acquisition}
\label{subsubsec:FLIM_acquisition}

Mapping of fluorescence lifetime was performed throughout the drying process, for both suspended liquid films and sessile droplets. In both case, initial time $t=0$ was set at the creation of the film or at the deposition of the droplet: acquisition was launched a few seconds later after placing the system on the FLIM.

For suspended films, images were initially acquired at a rate of $\SI{2}{\per\minute}$ during $\SI{90}{\minute}$, then at a rate of $\SI{0.1}{\per\minute}$ during the rest of the experiment, typically of a few hours and up to about one day. For sessile droplets, images were initially acquired at a rate of $\SI{2}{\per\minute}$ during $\SI{6}{\hour}$, then at a rate of $\SI{0.2}{\per\minute}$ until $\SI{21}{\hour}$ of experiment. In Fig.~\ref{fig:goutte_visco_vs_radius}, ~\ref{fig:goutte_conc_vs_radius} and ~\ref{fig:goutte_conc_vs_time}, only a part of the acquired data is displayed for better visualization.

\subsubsection{Image analysis}
\label{subsubsec:FLIM_analysis}

Lifetime data were determined for each pixel using a homemade Matlab GUI application. For all experiments, a reference image was acquired with a droplet of fluorescein deposited on a microscope glass slide, placed at the same distance of the objective lens as the sample in the corresponding experiment. Also, a background image was acquired for further subtraction. On acquired pictures, some pixels were associated with inconsistent lifetime values (typically negative, or larger than $\SI{10}{\nano\second}$). This is usually associated with some dead pixels of the CCD camera, areas without fluid or with lifetime below resolution limits: they were thus removed from analysis. More explanations on signal analysis can be found in Appendix A and in Ref.~\cite{vanmunster_2005}.

For suspended films, as can be seen in Fig.~\ref{fig:film_picture} for pictures at low $1.25 \times$ magnification, lifetime was mostly homogeneous in the central part of the film. Images were then acquired in this region with a higher $20 \times$ magnification and lifetime was averaged.

For sessile droplets, as discussed below in Section~\ref{subsec:results_drop}, lifetime is heterogeneous but displayed an axial symmetry. The contact line of the droplet was detected by thresholding the images with Matlab software. As a circular shape was obtained, a circular fit was applied in order to retrieve the coordinate of the center of the droplet. From this center, radial profiles were thus extracted by performing angular averages.

\subsection{Molecular rotor calibration}
\label{subsec:calibration}

In order to calibrate lifetime response of the FMR to viscosity, its lifetime was measured in the different water/fructose/glycerol solutions, both with and without SDS, described in Section~\ref{subsec:solutions}, and viscosity of the solutions were measured with a rheometer. Such calibration should be performed whenever the chemical nature of the species of the mixture is changed.

\begin{figure*}[!t]
    \includegraphics[width=0.9\textwidth]{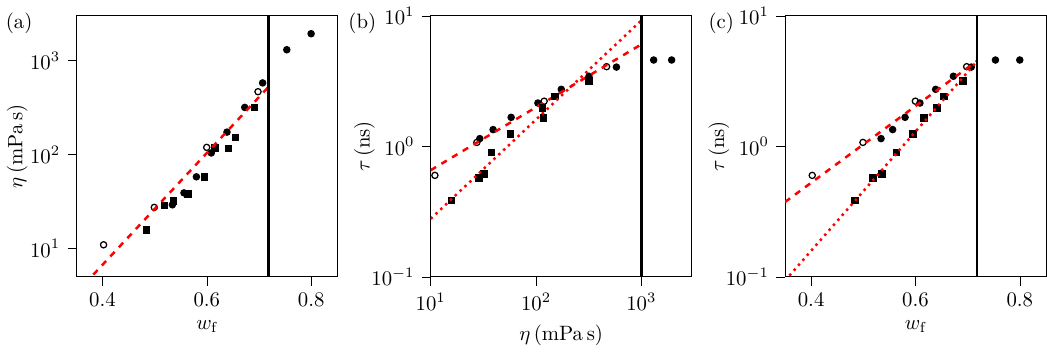}
    \caption{Calibration of the response of the rotor: viscosity $\eta$ and lifetime $\tau$ were measured in glycerol/water/fructose mixtures of different fructose mass fraction $w_\text{f}$. Squares ($\blacksquare$) correspond to solutions without SDS, used in experiments with sessile droplets. Circles correspond to solutions with a small amount of SDS, used in experiments with suspended films: filled symbols ($ \bullet$) correspond to an initial solution of fructose and glycerol with $w_\text{f} = 0.7$ and $w_\text{g} = 0.1$ progressively diluted with water in order to mirror the evaporation process, while open symbols ($\circ$) have a constant mass fraction of glycerol $w_\text{g} = 0.1$. Vertical lines delimit solutions saturated in fructose. (a) Evolution of viscosity $\eta$ with mass fraction of fructose $w_\text{f}$. The red dashed line represent a fit by Eq.~\eqref{eq:visco_mass_fraction} with $\eta_0 = \SI{2.8e-2}{\milli\pascal\second}$ and $w_0=\SI{7.3e-2}{}$. (b) Evolution of lifetime $\tau$ with viscosity $\eta$. The red lines represent fits along Eq.~\eqref{eq:Forster_Hoffmann}: dashed line is for solutions with SDS ($C = -0.66$ and $x=0.48$ if $\tau$ is expressed in $\SI{}{\nano\second}$ and $\eta$ in $\SI{}{\milli\pascal\second}$) and dotted line is for solutions without SDS ($C=-1.31$ and $x=0.76$ if $\tau$ is expressed in $\SI{}{\nano\second}$ and $\eta$ in $\SI{}{\milli\pascal\second}$).  (c) Evolution of lifetime $\tau$ with mass fraction of fructose $w_\text{f}$. The red lines represent fits along Eq. ~\eqref{eq:tau_mass_fraction}: dashed line is for solutions with SDS ($C' = -1.4$ and $w'_0=0.35$) and dotted line is for solutions without SDS ($C' = -2.5$ and $w'_0=0.22$).}
    \label{fig:calibration}
\end{figure*}

\subsubsection{Rheometry}
\label{subsubsec:rheo}

Rheological tests were conducted with a stress-controlled rheometer (Netzsch Kinexus) at a constant temperature of $\SI{20}{\celsius}$. According to the viscosity of the solution to be measured, a roughened cone-plate geometry ($\SI{4}{\degree} / \SI{40}{\milli\meter}$) or a double gap Couette geometry (inner radius $\SI{24}{\milli\meter}$, outer radius $\SI{27}{\milli\meter}$) was used. In both cases, a shear rate ramp was applied (from $\SI{1}{\per\second}$ to $\SI{10}{\per\second}$ in cone-plate geometry; from $\SI{50}{\per\second}$ to $\SI{5}{\per\second}$ in Couette geometry). The studied solutions were newtonian, and averaged viscosities are reported in the following paragraphs.

Evolution of the viscosity $\eta$ of the solutions with the mass fraction $w_\text{f}$ of fructose is displayed on Fig.~\ref{fig:calibration}(a). The results are similar for solutions containing surfactant (circle symbols) or not (square symbols), showing that the small amounts of SDS have no significant effect on the viscosity of the mixture.  

Two batches of solutions were prepared: in the first batch (open circles), mass fraction of glycerol was kept constant at $w_\text{g}=0.1$; in the second batch (filled circles), an initial solution (mass fractions of glycerol $w_\text{g}=0.1$ and of fructose $w_\text{f} = 0.7$) was progressively diluted by adding water, mirroring the evaporation that will be considered in later experiments. Similar lifetimes and viscosities were obtained in both cases. However, the measured viscosities were slightly higher than these reported for water/fructose mixtures~\cite{telis_2007}. This shows that while glycerol has an impact on the properties of the mixture, its changes of mass fraction are too small to affect viscosity of the solution and response of the FMR.

The evolution of viscosity $\eta$ with fructose mass fraction $w_\text{f}$ can be modeled along an exponential law:
\begin{equation}
    \eta = \eta_0 e^{w_\text{f} / w_0},
    \label{eq:visco_mass_fraction}
\end{equation}
\noindent with $\eta_0 = \SI{2.3e-2}{\milli\pascal\second}$ and $w_0=\SI{7.3e-2}{}$. Such an exponential evolution has been commonly used to describe viscosity of mixtures of carbohydrates with water~\cite{telis_2007}.

\subsubsection{Calibration}
\label{subsubsec:forster_hoffmann}

Drops of solution were deposited on a glass slide and lifetime $\tau$ of BODIPY-2-OH was measured as previously described for the different solutions, containing SDS or not. Its evolution with viscosity $\eta$ is represented in Fig.~\ref{fig:calibration}(b). Lifetime increased with viscosity as expected, and followed a power law known as F{\"o}rster-Hoffmann equation:
\begin{equation}
    \log \tau = C + x \log \eta,
    \label{eq:Forster_Hoffmann}
\end{equation}
\noindent where $C$ is a constant and $x$ is a dye-dependent exponent. When $\tau$ is expressed in $\SI{}{\nano\second}$ and $\eta$ in $\SI{}{\milli\pascal\second}$, the obtained values are respectively $C = -0.66$ and $x=0.48$ in presence of SDS (circle symbols) and  $C=-1.31$ and $x=0.76$ without SDS (square symbols).

The constant $C$ has no physical meaning but exponent $x$ is independent on the concentration of the mixture and quantifies the sensitivity of the FMR to viscosity. Such a law has already been reported for the same rotor in glycerol/DMSO mixtures with exponent $x=0.6$~\cite{nalatamby_2023}. While it can be justified on physical basis~\cite{forster_1971,loutfy_1986}, it should rather be seen as phenomenological and always experimentally validated before any use. It is indeed generally valid only on a certain range of viscosity values~\cite{dent_2015}. 

The value of the exponent $x$ is also slightly sensitive to the nature of the species in the mixture~\cite{dent_2015,li_2016,miao_2019,bittermann_2021a,briole_2021} and more generally, on the environment of the FMR~\cite{nalatamby_2023}. This could be an explanation of the different values obtained in solutions with or without SDS. We can hypothesize that the surfactant preferentially surrounds FMR molecules, thus hindering rotational motion (corresponding to an increase of lifetime) and decreasing the sensitivity to microviscosity (corresponding to a decrease of exponent $x$). Such a claim would require further work for being validated. In the following, the value of $x$ was obtained and used for each one of the studied solutions.

Finally, by comparing data (open and closed symbols), we can see that for the considered mixtures, the mass fraction of glycerol has little effect on the obtained results, consequently, viscosity and thus lifetime of the FMR are essentially determined by the fraction of fructose $w_\text{f}$. Mass fraction could thus be obtained from our measurements by combining Eq.~\eqref{eq:visco_mass_fraction} and Eq.~\eqref{eq:Forster_Hoffmann} along :
\begin{equation}
    \log \tau = C' + \frac{w_\text{f}}{w'_0}
    \label{eq:tau_mass_fraction}
\end{equation}
\noindent where $C' = C + x \log \eta_0$ and $w'_0 = \ln (10) \cdot w_0 / x$.

\subsection{Experimental device for suspended liquid films}
\label{subsec:exp_cell}

An experimental cell was designed to support the liquid film and control its environment. The cell was a square box of inner volume around $\SI{300}{\centi\meter\cubed}$ (width $\SI{5}{\centi\meter}$, height $\SI{6}{\centi\meter}$, depth $\SI{10}{\centi\meter}$), 3D printed with PLA filament (Raise 3D Premium, Black PLA, $\SI{1.75}{\milli\meter}$). The bottom half of the box was dedicated to the suspended film. The top half of the box contained a perforated shelf to hold a container of $\SI{8}{\milli\liter}$ for pouring a saturated aqueous salt solution, in order to stabilize humidity $\mathrm{RH}_\infty$ in the box, away from the film. Saturated solutions of sodium chloride $\mathrm{NaCl}$ (of water activity $a_\text{w} = 0.755$ at  $\SI{20}{\celsius}$) and of magnesium chloride $\mathrm{MgCl_2}$ (of water activity $a_\text{w} = 0.331$ at $\SI{20}{\celsius}$) were used~\cite{Fontana2007}.

Top and bottom of the box were closed using glass slides, in order to allow for optical access to the film and to the inside of the box during the evolution.

\begin{figure}[!]
    \includegraphics[width=0.8\columnwidth]{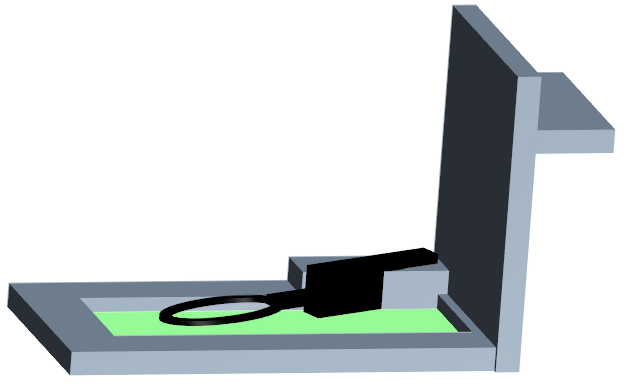}
    \caption{Schematics of the front door of the experimental cell. The liquid film is formed by dipping an elliptical frame (in black) in the solution. The frame is attached to a plain block, which is inserted in the door (in gray) to close the box after film formation. Optical access to the film is ensured by a microscope slide (in green) below.}
    \label{fig:film_support}
\end{figure}

In order to support the liquid film, an elliptical frame (of area $\mathcal{S} = \SI{204.2}{\milli\meter^2}$ and thickness $\SI{1}{\milli\meter}$) was also 3D printed in PLA. This frame was attached to a plain block of PLA (volume $\SI{1500}{\milli\meter^3}$): after film formation, this piece was inserted in the front door of the cell, in order to ensure horizontality of the film as displayed on Fig.~\ref{fig:film_support}.

For every experiment, the box was initially closed with a mock door and left to equilibrate in presence of the salt solution for about $\SI{20}{\minute}$. At $t=0$, the frame was dipped in the liquid solution to form the film, inserted in the door, which was then switched with the mock door to seal the box, trying not to disturb the air inside the box. The primary interest of the experimental cell was to protect the film from outside air flow in the room. Even with these precautions, relative humidity was never imposed at its expected value in the box, probably due to exchanges of vapor with the outside. However, presence of salt solutions still allowed to stabilize humidity in the box over a few hours, which was not possible else.
 
\section{Results}
\label{sec:results}

In the following paragraphs, mapping of lifetimes was obtained by using the FLIM and converted to obtain a mapping of viscosity through the Förster-Hoffmann equation~\eqref{eq:Forster_Hoffmann} with the appropriate calibration coefficients.

\subsection{Evaporation of liquid films}
\label{subsec:results_film}

In this Section, the presented results correspond to a solution of initial fructose and glycerol mass fractions of $w^0_\text{f} = 0.53$ and $w^0_\text{g} = 0.06$, with a saturated solution of sodium chloride to stabilize humidity.

\begin{figure*}[!]
    \includegraphics[width=0.7\textwidth]{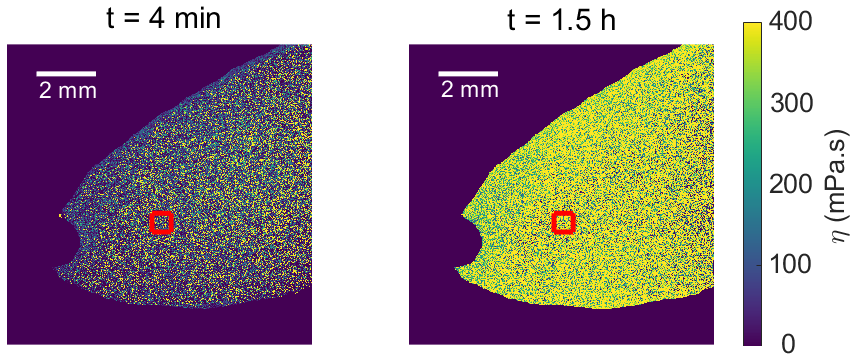}
    \caption{Viscosity maps of a large part of the liquid film observed at a low magnification $1.25 \times$, quickly after film formation (left) and after $\SI{1.5}{\hour}$ (right). Initial fructose and glycerol mass fractions were $w^0_\text{f} = 0.53$ and $w^0_\text{g} = 0.06$. A saturated $\mathrm{NaCl}$ solution was used to stabilize humidity away from the film. Viscosity appears to fluctuate homogeneously across the film. The red squares represent the part of the film which was observed with magnification $20 \times$ for further analysis, of dimension $\SI{630}{\micro\meter} \times \SI{640}{\micro\meter}$.}
    \label{fig:film_picture}
\end{figure*}

Figure~\ref{fig:film_picture} represents two mappings of viscosity obtained on a large part of the liquid film, with a magnification of $1.25 \times$, just after film formation and at a long time. Except from the edges of the film, viscosity appears as homogeneous on most of the film. Consequently, analysis was performed on a small region of dimension $\SI{630}{\micro\meter} \times \SI{640}{\micro\meter}$ in the middle of the film, using an objective lens of magnification $20 \times$ to obtain more precise lifetime measurements. In this case, the depth over which lifetime was averaged was smaller than the thickness of the film: no systematic variation of lifetime was observed when changing the focus of the microscope which shows there were no significant variation of fructose concentration in the direction perpendicular to the film.

\begin{figure}[!]
    \includegraphics[width=0.8\columnwidth]{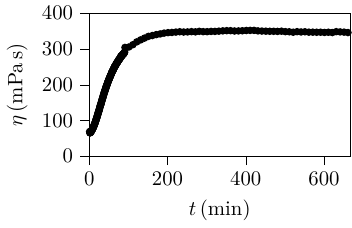}
    \caption{Evolution of averaged viscosity on the observation zone, in the middle of the film with a $20 \times$ objective lens. Initial fructose and glycerol mass fractions were $w^0_\text{f} = 0.53$ and $w^0_\text{g} = 0.06$. A saturated $\mathrm{NaCl}$ solution was used to stabilize humidity away from the film.}
    \label{fig:film_visco_example}
\end{figure}

As viscosity was homogeneous across the film, it was averaged on the region of observation: its evolution with time is displayed on Fig.~\ref{fig:film_visco_example}.

Viscosity progressively increased over a few hours, before reaching a steady state persisting over hours, provided humidity in the box did not fluctuate. Qualitatively, such an increase of viscosity can be attributed to the slow evaporation of the film: while water evaporates from the liquid, it progressively gets concentrated in fructose and glycerol. For preliminary experiments in open air (i.e. without the box, not detailed here), at long times, film finally turned into a sticky crust of fructose and glycerol, from which most of water disappeared.

Experiments were performed using different initial fructose mass fractions, and different saturated salt solutions ($\mathrm{NaCl}$ or $\mathrm{MgCl_2}$) to stabilize humidity and repeated two to three times for all conditions. While this qualitative overall behavior was reproducible from one experiment to another and remained valid whatever the conditions, quantitative values displayed noticeable variability, even when keeping similar conditions. This is most likely due to the limited repeatability of the film formation by dipping, in particular as the initial thickness of the film changed. 

In order to estimate this effect, weighing experiments were performed to determine mass change when bursting the film. For the solution presented here for instance, a mass of $m_\text{film} = \SI{54(10)}{\milli\gram}$ was obtained by weighing three films, corresponding to a film thickness $e_\text{film} = m_\text{film} / \rho_\text{sol} \mathcal{S} = \SI{220(40)}{\micro\meter}$ where $\rho_\text{sol} \simeq \SI{1.2e3}{\kilo\gram\per\meter\cubed}$ was the density of the solution, which was measured by weighing a controlled volume of solution and was consistent with density values obtained from literature for fructose/water solutions~\cite{darrosbarbosa_2003,fucaloro_2007}. By performing such measurements with different mass fractions of fructose, a trend appeared of increasing film thickness with fructose fraction.

Changes of initial composition of the liquid mixture induced consistent evolution: mixtures of higher fructose fraction led to initially larger viscosities, and smaller range of variation, as the final concentration was determined by the external humidity and did not depend on the initial solution content. It is not clear whether the initial fructose fraction influences film dynamics in itself: no systematic effect was observed, larger than fluctuation when reproducing the experiments.

Changing the salt solution did also not lead to a systematic change. Consequently, the experimental cell was probably not efficient enough to isolate the inner atmosphere from outside humidity. Salt solution allowed to stabilize humidity on rather long times (of a few hours) but for longer duration (of a few ten hours), fluctuations were observed, probably associated with changes of humidity in the room. Results for viscosity evolution using different initial composition and salt solutions are reported in more details in Appendix E, in Figure~\ref{fig:visco_variousCI_SI_tot}).

\subsection{Evaporation of a sessile droplet}
\label{subsec:results_drop}

\begin{figure*}[!]
    \includegraphics[width=0.9\textwidth]{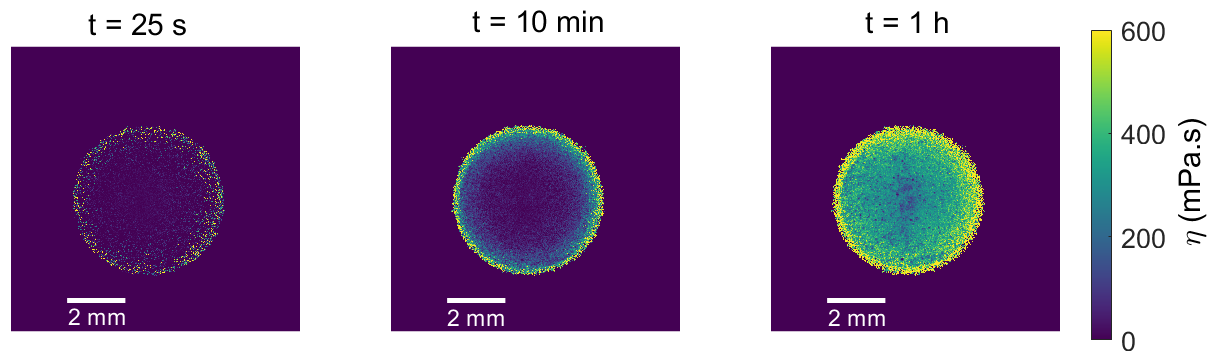}
    \caption{Viscosity map of a sessile droplet of respective initial mass fractions of fructose and glycerol $w^0_\text{f} = 0.48$ and $w^0_\text{g} = 0.06$, deposited on a glass slide and evolving to open air. The progressive accumulation in time of fructose at the sides of the drop due to evaporation-induced flow can be observed, as well as the progression of the concentration gradient and the overall increase of viscosity.}
    \label{fig:goutte_picture}
\end{figure*}

Contrary to the case of the liquid film, evaporation of sessile droplets lead to heterogeneous viscosity profiles, as can be seen in Fig.~\ref{fig:goutte_picture}. The results described here were obtained by depositing on a glass slide a droplet of solution of initial mass fractions of fructose and glycerol $w^0_\text{f} = 0.48$ and $w^0_\text{g} = 0.06$. For these experiments, performed with a $1.25 \times$ objective lens, the depth over which lifetime was averaged was larger than the thickness of the droplet: possible variation of concentration in the direction of light propagation thus could not be measured.

\begin{figure}[!]
    \includegraphics[width=0.8\columnwidth]{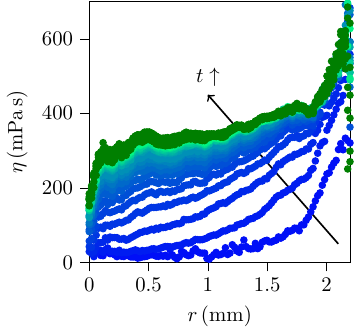}
    \caption{Viscosity profile within the droplet (angular average) at different times, with respective initial mass fractions of fructose and glycerol $w^0_\text{f} = 0.48$ and $w^0_\text{g} = 0.06$. Time is color coded: graded colors at early times are every $\SI{5}{\minute}$ from $\SI{6}{\minute}$ to $\SI{2}{\hour}$, then there is no significant evolution and curves are displayed in dark green, every hour for times between $\SI{2}{\hour}$ and $\SI{21}{\hour}$.}
    \label{fig:goutte_visco_vs_radius}
\end{figure}

No particular surface treatment was performed on the supporting glass slide and drying essentially occurred at constant drop radius mode, as the contact line remained pinned. However, viscosity profiles remained mostly axisymmetric: consequently, angular averages were performed to obtain viscosity profiles as a function of the distance $r$ to the center of the drop as explained in Section~\ref{subsubsec:FLIM_analysis}. The obtained viscosity profiles are displayed on Fig.~\ref{fig:goutte_visco_vs_radius}.

Overall, lifetime in the droplet increased over time, corresponding to an increase of fructose and glycerol concentration due to water evaporation, similar to the case of liquid films. However, viscosity was not homogeneous: very soon after the deposition of the droplet, a gradient of viscosity developed at the edge of the drop, and the initial homogeneous state was not observed in our experiment, due to the initial time (about one minute) required to prepare the setup and launch the acquisition. Within a few minutes, this gradient developed across the whole droplet. Then, it got progressively smoothed until reaching a quasi-steady state after a few hours, in which viscosity was almost homogeneous across the droplet. At this stage, the mixture was close to saturation so fructose crystallized, while solution kept a constant concentration. Rotors were then only measuring the constant viscosity of the saturated solution; this does not correspond to full drying of the droplet as water kept evaporating.

In the center of the droplet, a gradient of viscosity over short distances (about $\SI{100}{\micro\meter}$) appeared to persist, which is due to imperfect determination of the center of the droplet and to the presence of some small air bubbles at this position on the presented data set. Also, viscosity strongly increased at the edge of the droplet, which is related to the ``coffee-ring effect'': fructose accumulated there and crystallized, leading to an increased apparent viscosity. This region is not analyzed in what follows because rotor was there trapped in the crust and the presented calibration of its response was not reliable in these conditions.

\begin{figure}[!]
    \includegraphics[width=0.8\columnwidth]{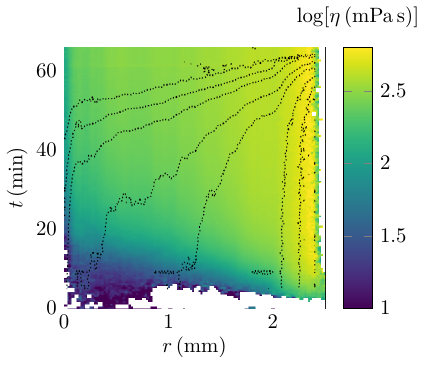}
    \caption{Spatio-temporal representation of the radially averaged viscosity profile of the droplet of respective initial mass fractions of fructose and glycerol $w^0_\text{f} = 0.48$ and $w^0_\text{g} = 0.06$, with viscosity in logarithmic scale. White spaces for early times or large distance correspond to inconsistent lifetime data, either as initial viscosity is too small and associated lifetimes are below the resolution range of the FLIM, or as there is no measurable signal beyond the droplet radius. Dashed lines are contour plots corresponding to $\log \eta \, [\SI{}{\milli\pascal\second}]= 1.7$, $2$, $2.2$, $2.4$, $2.5$, $2.6$ and $2.7$ (from left to right).}
    \label{fig:spatio_temp_logvisco_goutte}
\end{figure}

Finally, evolution of viscosity field can be represented on the spatio-temporal diagram of Fig.~\ref{fig:spatio_temp_logvisco_goutte}. Viscosity is color coded in logarithmic scale. Contour lines are represented to facilitate reading. Some white pixels appeared, which were related to inconsistent lifetime values as discussed in Section~\ref{subsubsec:FLIM_analysis}, either due to lifetime below resolution limit at short times or at large radius, out of the droplet.

\section{Discussion}
\label{sec:discussion}

Results presented in the previous section show that FMR and FLIM can be used to map viscosity in confined systems evolving in time, with observations consistent with established phenomenology of drying process. To go beyond these observations, further analysis can be performed on these measurements. As the evaporation dynamics is influenced by concentration field rather than viscosity, in what follows, measured lifetimes were directly converted into fructose mass fraction $w_\text{f}$ using Eq.~\eqref{eq:tau_mass_fraction}. 

\subsection{Evaporation of liquid films}
\label{subsec:discussion_film}

As was mentioned in Section~\ref{subsec:results_film}, evaporation of the liquid film occurs at almost homogeneous viscosity, thus concentration, in the film. Consequently, only little flow occured in the film and the drying process can be modeled as unidirectional. In the presence of an unsaturated atmosphere, an evaporative flux of water is driven by the unbalance between water activity in the gas phase (which is related to the relative humidity $\mathrm{RH}$) and in the liquid phase. Evaporation in turn progressively concentrates the non volatile solutes and increases water activity, until an equilibrium is eventually reached or until fructose crystallizes and water dries out (case not considered in the present experiments). 

\begin{figure}[!]
    \includegraphics[width=0.8\columnwidth]{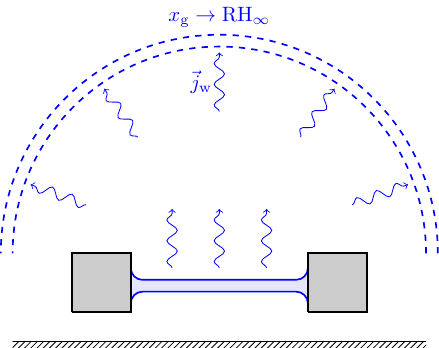}
    \caption{Schematics of the liquid film, supported by the frame (in gray). Water vapor flux $\vec{j}_\text{w}$ is represented in blue: it is driven by the humidity $\mathrm{RH}_\infty$ far away from the film. As the film is small enough, evaporative flux is considered to be similar to this that would exist in an open atmosphere. As water evaporates, non volatile solutes (fructose and glycerol) get concentrated, which increases water activity and slows down evaporation.}
    \label{fig:schema_film}
\end{figure}

The film had a typical extension of about $\SI{1}{\centi\meter}$, which is small compared to box lateral dimensions of a few centimeters: in first approximation, the film can be assumed to be placed in an open atmosphere, with humidity imposed at infinity. A schematic of the model situation is given on Fig.~\ref{fig:schema_film}.

Details on the model are provided in Appendix C. Briefly, it adapts the model proposed by Dollet and Boulogne~\cite{dollet_2017} for evaporation of a circular water pool, to account for the effect of non volatile solutes. In the conditions of the experiment, water flux in the gas phase is mostly caused by natural convection. This flux in turn imposes the evolution of concentrations in the liquid film. By assuming that the mixture behaves ideally, the following differential equation can be obtained for the mole fraction of water $x_\text{w}$:
\begin{equation}
    \frac{1}{(1-x_\text{w})^2} \frac{\mathrm{d} x_\text{w}}{\mathrm{d} t} = - \frac{x_\text{w} - \mathrm{RH}_\infty}{\tilde{\tau}}  \left[A_1 \left(\frac{x_\text{w} - \mathrm{RH}_\infty}{\mathrm{RH}_\infty}\right)^{1/5} + a_2\right]. 
    \label{eq:model_evap_film}
\end{equation}
\noindent where $A_1$ and $a_2$ are parameters determined in Ref.~\cite{dollet_2017} and $\tilde{\tau}$ is a characteristic time, related to various constants and proportional to the amount of non volatile solutes in the liquid, which was constant during the whole evolution.

\begin{figure}[!]
    \includegraphics[width=0.8\columnwidth]{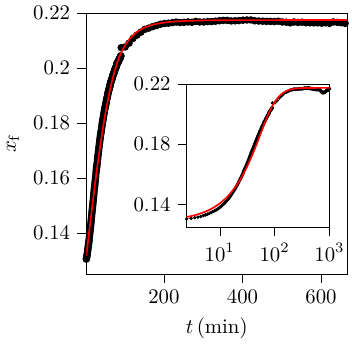}
    \caption{Evolution of the mole fraction of fructose in the film over time, for a film of initial fructose and glycerol mass fractions $w^0_\text{f} = 0.53$ and $w^0_\text{g} = 0.06$ (corresponding to a mass ratio $\alpha_w = 0.12$). Red line represents prediction by model~\ref{eq:model_evap_film}, with $\mathrm{RH}_\infty = 0.731$ estimated from steady state, and $\tilde{\tau} = \SI{3.8}{\minute}$ obtained by curve fitting. Inset represents the same data with logarithmic time scale.}
    \label{fig:film_fit_example}
\end{figure}

In order to compare the model with experimental data, mole fractions of fructose and water, $x_\text{f}$ and $x_\text{w}$, should first be obtained from measurement of fructose mass fraction $w_\text{f}$. This can be done by exploiting the fact that the glycerol to fructose mass ($\alpha_w = w^0_\text{g}/w^0_\text{f}$) and mole ($\alpha_x = (M_\text{f}/M_\text{g}) \alpha_w$) ratios are constant during evaporation, set by the initial composition of the mixture, as detailed in Appendix B. In the analyzed experiment, $\alpha_w = 0.12$ and $\alpha_x = 0.24$. Evolution of fructose mole fraction $x_\text{f}$ is displayed on  Fig.~\ref{fig:film_fit_example}. 

Relative humidity away from the film was estimated from the mole fraction at steady state $x^\infty_\text{f}$: it is indeed equal to the mole fraction of water in the film at steady state so $\mathrm{RH}_\infty = 1 - (1+\alpha_x)x^\infty_\text{f}$. Once this parameter was known, a solution of Eq.~\eqref{eq:model_evap_film} could be numerically obtained as function of parameter $t/\tilde{\tau}$ (using the solve\_ivp function from Python library Scipy-Integrate). Finally, a least-square principle was applied (using the curve\_fit function from Python library Scipy-Optimize) to determine the optimal value of $\tilde{\tau}$, best matching data with model solution. Corresponding results are represented on Fig.~\ref{fig:film_fit_example}.

Agreement between experimental data and model appears as satisfactory even if not perfect, as emphasized on the logarithmic plot in the inset of Fig.~\ref{fig:film_fit_example}. The obtained value of $\tilde{\tau} = \SI{3.8}{\minute}$ allows to estimate an initial thickness of the film $e_\text{film} \simeq \SI{400}{\micro\meter}$. This value is larger, yet of similar order of magnitude than this estimated by weighing the film, described in Section~\ref{subsec:results_film}.

Better agreement between data and model can be obtained. In particular, parameters $A_1$ and $a_2$ were set to the values obtained experimentally in Ref.~\cite{dollet_2017} but they could be used as further fitting parameters, and modify noticeably the estimated initial film thickness.

However, it is important to acknowledge that such model is not pretended to be quantitative. Geometry of the problem is simplified: effect of the meniscus on drying dynamics is neglected, elliptical shape of the film is approximated by an equivalent circular shape, evaporation from below the film is neglected and the film is assumed to be in an open space with humidity imposed at infinity. While these assumptions are debatable, a more realistic model would include geometrical corrections which could modify the expression of $\tilde{\tau}$ but should not change qualitatively the predictions. 

More importantly, assuming ideality of the mixture is only a rough assumption (see Appendix D) in binary water/fructose and water/glycerol mixture which is not guaranteed to remain valid in the ternary mixture: such non-ideality could induce further nonlinearity in Eq.~\ref{eq:model_evap_film} and change the dynamics predicted by the model. 

Consequently, the model is expected to be qualitative only. The fact that it describes satisfactorily the evaporation dynamics and that it allows to retrieve an acceptable order of magnitude of the initial film thickness validates that it captures the essential features of the phenomenon. It is to note that, by considering films of different initial fructose fractions, the observed trend of increasing film thickness observed when weighing the film is not reproduced by this analysis: however, a more systematic study of this initial thickness should be done to reach a clear conclusion. A similar conclusion remains valid for experiments with different initial composition or external humidity, as reported in Appendix E (in particular in Table~\ref{tab:table1}).

Finally, the qualitative success of this simple model shows that it captures the essential phenomena at play in the evaporation process. It also validates the proposed methodology of viscosity measurement using FMR, to retrieve useful, quantitative information on suspended liquid film.

\subsection{Evaporation of a sessile droplet}
\label{subsec:discussion_drop}

\begin{figure*}
    \includegraphics[width=0.8\textwidth]{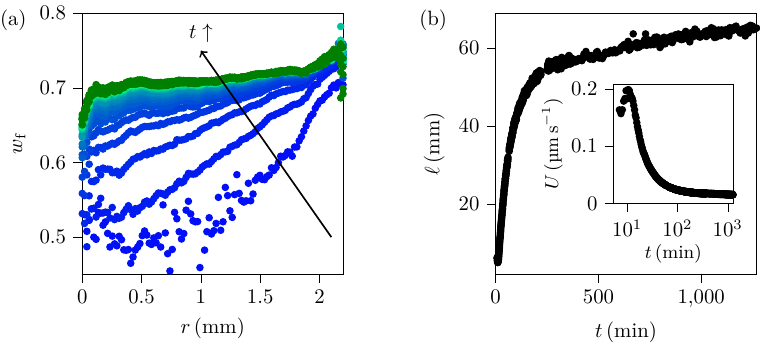}
    \caption{Analysis of concentration profile during evaporation in open air of a sessile droplet at different time. Initial mass fractions of fructose and glycerol were respectively $w^0_\text{f} = 0.48$ and $w^0_\text{g} = 0.06$. (a) Fructose mass fraction profile, obtained from angular average of lifetime through~\eqref{eq:tau_mass_fraction}. Profile is represented at different time which are color coded.Curves with color code are separated by $\SI{5}{\minute}$ from acquisition start to $\SI{100}{\minute}$, then steady state is reached and curved are represented in green, separated by $\SI{1}{\hour}$ until $\SI{21}{\hour}$ of acquisition. (b) Evolution of the slope $\ell$ of mass fraction profile in the middle of the drop. Inset : Characteristic flow speed in the droplet $U = D/\ell$ as a function of time in logarithmic scale.}
    \label{fig:goutte_conc_vs_radius}
\end{figure*}

As discussed in the Introduction, evaporation of a droplet of a pure fluid is by itself a very rich and complex phenomenon: in the considered situation of evaporation of ternary droplet, a quantitative modeling of the evolution of concentration is thus completely out of reach, and would require detailed thermodynamics modeling coupled to hydrodynamics simulations. However, some quantitative characteristics of the system can be retrieved by further analyzing viscosity maps, that can be converted in profiles of fructose mass fraction within the droplet using Eq.~\eqref{eq:visco_mass_fraction}.

Evolution of mass fraction field $w_\text{f}$ in the droplet over time is represented on Fig.~\ref{fig:goutte_conc_vs_radius}(a) from the results discussed in Section~\ref{subsec:results_drop}. It is qualitatively similar to this of viscosity profile, with an overall increase of concentration associated with the initial build up, then relaxation of a gradient over the first hour.

In the intermediate region of the droplet, for $\SI{0.1}{\milli\meter} < r < \SI{1.7}{\milli\meter}$ where lifetime measurements are significant, concentration profiles were roughly linear. It was thus possible to perform linear fits in order to extract a time-dependent characteristic length for concentration profile $\ell$ within the droplet, as pictured in Fig.~\ref{fig:goutte_conc_vs_radius}(b). It is important to note that this analysis does not rely on any physical model: rather, it is a way to define a typical length scale. Measured values of $\ell$ were of the order or larger than the drop size, which was consistent with the small variations of concentration.

This length can be seen as resulting from the competition between advection by the inner flow, driven by evaporation, which builds the concentration gradient by accumulating fructose at the edge of the droplet, and diffusion which tends to homogeneize the concentration. By considering a typical diffusion coefficient $D \sim \SI{e-9}{\meter\squared\per\second}$, the average flow speed $U$ required to build a concentration gradient on length $\ell$ is given by $U \sim D/\ell$ and is represented in the inset of Fig.~\ref{fig:goutte_conc_vs_radius}(b). It is of the order of $\SI{0.1}{\micro\meter\per\second}$ and decays over time, due to the decay of evaporative flux as water activity in the solution approaching this of the ambient atmosphere.

\begin{figure*}[!]
    \includegraphics[width=0.9\textwidth]{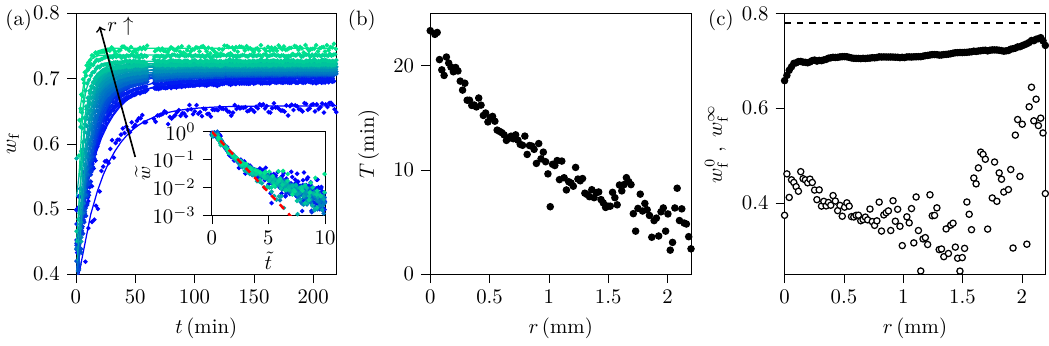}
    \caption{Analysis of evolution of concentration during evaporation in open air of a sessile droplet at different positions. Initial mass fractions of fructose and glycerol were respectively $w^0_\text{f} = 0.48$ and $w^0_\text{g} = 0.06$. (a) Evolution of fructose mass fraction, obtained from angular average of lifetime through~\eqref{eq:tau_mass_fraction}, at different positions in the drop. Distance to the center is color coded: successive curves correspond to positions separated by $\SI{100}{\micro\meter}$. Continuous line represents fitting along Eq.~\eqref{eq:drop_mass_fraction_time_fit}. Inset : Normalized fructose fraction $\tilde{w} = (w_\text{f} - w^0_\text{f}) / (w^\infty_\text{f} - w^0_\text{f})$ as a function of normalized time $\tilde{t} = t/T$ in semilogarithmic scale. Perfect agreement with the model would correspond to point placed on the red dashed line. (b) and (c) Respective evolution of the characteristic $T$ and of the initial $w^0_\text{f}$ ($\circ$) and steady $w^\infty_\text{f}$ ($\bullet$) mass fraction in the droplet with the distance $r$, obtained by fitting evolution along Eq.~\eqref{eq:drop_mass_fraction_time_fit}.} 
    \label{fig:goutte_conc_vs_time}
\end{figure*}

Alternatively, time evolution of fructose fraction can be considered at different points of the droplet, as displayed on Fig.~\ref{fig:goutte_conc_vs_time}(a). Everywhere in the droplet, concentration progressively increases as water evaporates, until reaching a quasi-steady state with a persistent weak gradient. It appears that points at the periphery of the droplet have a quicker dynamics than in the middle.

Such an effect can be further quantified by fitting fructose fraction evolution with a phenomenological exponential model:
\begin{equation}
    w_\text{f} (r,t) = w^0_\text{f} (r) + (w^\infty_\text{f} (r) - w^0_\text{f} (r)) (1-e^{-t/T(r)})
    \label{eq:drop_mass_fraction_time_fit}
\end{equation}
\noindent where $w^0_\text{f} (r)$, $w^\infty_\text{f} (r)$ and $T(r)$ are respectively the (position-dependent) initial mass fraction, final mass fraction, and characteristic time. Again, the exponential fit is only a way to define a characteristic time and does not rely on a physical model: as can be seen from the mastercurve in the inset of Fig.~\ref{fig:goutte_conc_vs_time}, it describes satisfactorily experimental data for time up to about $5 T$.

Resulting characteristic time $T(r)$ is represented in Fig.~\ref{fig:goutte_conc_vs_time}(b): it is of the order of a few minutes and decreases when approaching the edge of the droplet. This is consistent with the increase of the evaporative flux when approaching the triple line, due to geometric singularity~\cite{guena_2007}. The initial mass fraction $w^0_\text{f} (r)$ appears as roughly constant in the middle of the droplet, with an increase on the edges as concentration gradient started to build on when acquisition was launched. Finally, final mass fraction $w^\infty_\text{f} (r)$ softly increases, and is smaller but of the order of the estimated saturation point of the ternary water/fructose/glycerol mixture.

\section{Conclusion}
\label{sec:conclusion}

In this study, a coupled use of Fluorescent Molecular Rotors and Fluorescence Lifetime Imaging Microscopy was introduced as a local, in-situ tool to monitor evolution of physico-chemical transformations in confined systems, exemplified on the important case of evaporation and drying of liquid mixtures involved in numerous material formulation processes. In such phenomenon, unbalance of solvent activity in the liquid and in the gas phase leads to solvent evaporation: in turn, solutes get concentrated and solvent activity progressively increases, slowing down (and possibly arresting) drying dynamics. Concentration fields are thus a crucial parameter for monitoring such processes. Sensitivity of lifetime to local viscosity allowed for a local mapping of viscosity, which could be related to concentration fields: the presented methodology can thus be adapted to any process involving changes of viscosity, which is often found in industrial systems.

First, the case of evaporation of a thin liquid film was considered: in this situation, drying was homogeneous and the situation could be modeled. While a full quantitative characterization of the phenomena was out of reach, in particular due to the presence of two non volatile solutes in the liquid, the simple model proposed here is in satisfactory agreement with experimental measurements, thus validating the methodology. Then, the more complex situation of evaporation of a sessile droplet on a solid substrate was studied. In this case, concentration fields were related to the competition between diffusion in the liquid phase and evaporation-driven flows. Such a complex process cannot be analytically modeled, but use of FMR and FLIM allowed to retrieve interesting parameters to characterize the dynamics, thus affording a new tool for studying drying experiments and controlling materials formulation involving evaporation steps.

The methodology presented here thus opens new paths for monitoring of physico-chemical transformations, involving changes of viscosity. The method is local, with resolution only limited by the employed objective lens, and poorly invasive as only little FMR concentration is required. The main limitation currently relies in the limited solubility of available FMR in aqueous media.

\acknowledgements{The authors are grateful to Javier Ord{\'o}{\~n}ez-Hern{\'a}ndez and Norberto Farf{\'a}n for providing the molecular rotor. They thank Jean-Baptiste Salmon for discussions on evaporation models and more generally on the experiments, and Adrien Martin for his participation in some experiments. They also thank Sophie Galinat for her help with attempts of UNIFAC modeling and navigation in the Dortmund Data Bank. The research presented here was funded by French ANR grant MicroVISCOTOR (ANR-18-CE42-0010-01). The authors also acknowledge Solvay and CNRS for support.

This document is the unedited Author's version of a Submitted Work that was subsequently accepted for publication in ACS Applied Materials and Interfaces, copyright 2024 American Chemical Society after peer review. To access the final edited and published work see \url{https://pubs.acs.org/doi/10.1021}. It is to note that in the published version, the number of references has been decreased to comply with journal's guidelines.}

%--------------------------------------------------------------------

\appendix

\section{Data analysis in Fluorescence Lifetime Imaging Microscopy}
\label{sec:app_FLIM}

The FLIM setup presented in this paper operates in frequency domain, as opposed to time domain measurements using photocounters, that directly measure fluorescence decay after a pulsed excitation. Here, excitation light is modulated harmonically at a frequency $f$ and steady state fluorescence response is measured. As the absolute phase of the excitation signal is unknown, analysis of the response is performed using a lock-in amplifier, with a reference signal synchronous to the excitation signal and tunable phase. By fitting the result obtained with different phases, it is possible to retrieve the modulation depth $M_\text{samp}$ and the phase shift $\varphi_\text{samp}$ of the fluorescent signal with respect to the excitation. In order to get rid of phase shift related to instrumental response and light propagation, a preliminary measurement of modulation depth $M_\text{ref}$ and the phase shift $\varphi_\text{ref}$  should be performed with a material of tabulated lifetime $\tau_\text{ref}$. Then, the ratio $M_\text{samp}/M_\text{ref}$ and the phase difference $\varphi_\text{samp}-\varphi_\text{ref}$ are directly related to the lifetime of the sample $\tau$. As the latter is more reliable, the lifetime $\tau$ is obtained from the phase delay through:
\begin{equation}
    \tau = \frac{1}{\omega} \tan\left[\phi_{\text{samp}} - \phi_{\text{ref}} + \arctan(\omega \tau_{\text{ref}})\right].
    \label{eq:lifetime-definition}
\end{equation}
\noindent where $\omega=2 \pi f$. FLIM setup allows to perform such an analysis pixel per pixel, generating a lifetime image of the observed sample.

\section{Conversion between mole and mass fractions}
\label{sec:app_conversion}

As explained in main text, experiments were performed using a ternary water/fructose/glycerol mixture. Lifetime measurements allowed to extract either viscosity $\eta$ or fructose mass fraction $w_\text{f}$. However, evaporation model for the suspended film requires the knowledge of water mole fraction $x_\text{w}$ to determine its activity, and analysis of data exploits fructose mole fraction $x_\text{f}$. 

Fructose mole fraction $x_\text{f}$ is defined by:
\begin{equation}
    x_\text{f} = \frac{(w_\text{f} / M_\text{f})}{(w_\text{f} / M_\text{f}) + (w_\text{g} / M_\text{g}) + (w_\text{w} / M_\text{w})}
\end{equation}
\noindent where $w_\text{f}$, $w_\text{g}$ and $w_\text{w}$ are the respective mass fractions of fructose, glycerol and water. As fructose and glycerol are non-volatile species, their mass ratio $\alpha_w = w_\text{g} / w_\text{f}$ is constant during the film evolution and determined by the initial composition of the mixture. Consequently, $w_\text{g} = \alpha_w w_\text{f}$ and $w_\text{w} = 1 - (1+\alpha_w) w_\text{f}$ can be determined from fructose mass fraction measurements, and mole fractions can eventually be computed.

Similarly, the glycerol to fructose mole ratio $\alpha_x = x_\text{g} / x_\text{f} = (M_\text{f} / M_\text{g}) \alpha_w$ is constant, which allows to obtain the different mole fractions from $x_\text{f}$.

\section{Detailed model for film evaporation}
\label{sec:app_model}

\subsection{Convective and diffusive fluxes}

As showed in Fig.~\ref{fig:film_picture} of the main text, concentration field is homogeneous across most of the liquid film during all the drying process. Moreover, as the frame supporting the film is close of the bottom of the box, it can be assumed that the space below the film is quickly with water vapor in equilibrium with the film, and that most of the evaporation occurs from the part above. Finally, as the liquid film (of typical radius $\SI{1}{\centi\meter}$) is smaller than the size of the box (of typical dimension of a few centimeters), the situation under study can be roughly assimilated to this of evaporation of a liquid pool in an infinite atmosphere, with imposed humidity at long distance. Such a problem was studied experimentally and modeled by Dollet and Boulogne~\cite{dollet_2017} for the case of a circular pool of radius $R$. As in the experiments presented here, the film is ellipsoidal, an effective radius $R = \sqrt{\mathcal{S}/\pi}$ will be considered to maintain a similar evaporating surface $\mathcal{S}$.

While evaporation is usually modeled as driven by diffusive flux in the vapor phase, contribution of natural convection should also be accounted for, as water fraction gradient induces a gradient of air density. Competition between the two fluxes is quantified by the dimensionless Grashof number:
\begin{equation}
    \mathrm{Gr}  = \frac{x^\text{g}_\text{w,0} - x^\text{g}_\mathrm{w,\infty}}{x^\text{g}_\mathrm{w,\infty}} \frac{gR^3}{\nu^2} 
\end{equation}
\noindent where $x^\text{g}_\text{w,0}$ and $x^\text{g}_\mathrm{w,\infty}$ are the mole fractions of water in gas phase respectively at the film level and far away from the film, $g \, [\SI{}{\meter\per\second\squared}]$ is the gravity acceleration and $\nu\, [\SI{}{\meter\squared\per\second}]$ is the kinematic viscosity of air. 

In the considered experiment, $\mathrm{Gr} \sim 4000$ so the convective contribution is dominant. Following analysis of Ref.~\cite{dollet_2017}, the evaporating mass flux of water $\Phi_\text{w} \, [\SI{}{\kilo\gram\per\second}]$ in such conditions can be written as:
\begin{equation}
    \Phi_\text{w} = 2\pi D \rho_\text{air} R \frac{M_\text{w}}{M_\text{air}}  (x^\text{g}_\text{w,0} - x^\text{g}_\mathrm{w,\infty}) (a_1\mathrm{Gr}^{1/5} + a_2)
\end{equation}
\noindent where $M_\text{w}$ and $M_\text{air} \, [\SI{}{\kilo\gram\per\mole}]$ are the molar mass of water and air. Finally, $a_1$ and $a_2$ are free dimensionless parameters, which are experimentally estimated to $a_1  =0.3$ and $a_2 = 0.5$. It is important to keep in mind that this rate does not correspond to a diffusive flux, as diffusion was neglected in the limit $\mathrm{Gr} \gg 1$. The term $a_1 \mathrm{Gr}^{1/5}$ corresponds to the flux in the middle of the film which is not linear in mole fraction gradient (as $\mathrm{Gr}$ depends on it), and the second term $a_2$, despite corresponding to a linear flux, also stems from convection at the edges of the film.

\subsection{Evaporation rate expressed as a function of liquid water activity}

Water mole fraction in air is related to activity $a_\text{w}$ through:
\begin{equation}
    x^\text{g}_\text{w} = \frac{P_\text{sat}^+}{P} e^{v_\text{w}^+ (P-P^+_\text{sat}) / \mathcal{R} \theta} a^\text{g}_\text{w}
\end{equation}
\noindent where $P_\text{sat}^+ \, [\SI{}{\kilo\gram\per\meter\cubed}]$ is the saturation pressure of pure water at the considered temperature $\theta \, [\SI{}{\kelvin}]$, $P \, [\SI{}{\pascal}]$ is the atmospheric pressure, $v_\text{w}^+$ is the molar volume of pure water, and $\mathcal{R}$ is the ideal gas constant. For the experiment presented here, the exponential Poynting correction is close to $1$. By assuming the vapor is ideal and that the vapor at film level is in equilibrium with the liquid, mole fraction of the gas at the film level is equal to the activity of water in the liquid film ($x^\text{g}_\text{w,0} = a_\text{w}$) and activity of water far from the liquid film coincides with relative humidity ($x^\text{g}_\mathrm{w,\infty} = \mathrm{RH_\infty}$). Finally, by introducing $K = P/P^+_\text{sat}$, the mass flux of water can be rewritten as
\begin{multline}
    \Phi_\text{w} = 2\pi D \rho_\text{air} R \frac{K M_\text{w}}{M_\text{air}}  (a_\text{w} - \mathrm{RH}_\infty) \times \\ \left[a_1\left( \frac{gR^3}{\nu^2}\right)^{1/5} \times \left(\frac{a_\text{w} - \mathrm{RH}_\infty}{\mathrm{RH}_\infty}\right)^{1/5} + a_2\right].
    \label{eq:model_evap_flux}
\end{multline}

\subsection{Evolution of water fraction in the film}

As shown in Eq.~\eqref{eq:model_evap_flux}, evaporative flux is driven by the difference in activity of water in the liquid film and far away in the gas phase: in turn, as water evaporates, the liquid gets concentrated in fructose and glycerol and water activity evolves. By definition,
\begin{equation}
    \Phi_\text{w} = - \frac{\mathrm{d} m_\text{w}}{\mathrm{d} t} = - M_\text{w} \frac{\mathrm{d} n_\text{w}}{\mathrm{d} t}
    \label{eq:model_flux_definition}
\end{equation}
\noindent where $m_\text{w} \, [\SI{}{\kilo\gram}]$ and $n_\text{w} \, [\SI{}{\mole}]$ are the mass and amount of water in the film. Water mole fraction $x_\text{w}$ in the film can be written as:
\begin{equation}
    x_\text{w} = \frac{n_\text{w}}{N + n_\text{w}}
\end{equation}
\noindent where $N$ is the (constant) amount of non volatile species (glycerol and fructose) in the film. Consequently,
\begin{equation}
    \frac{\mathrm{d} x_\text{w}}{\mathrm{d} t} = \frac{(1-x_\text{w})^2}{N} \frac{\mathrm{d} n_\text{w}}{\mathrm{d} t}.
    \label{eq:model_fraction_derivative}
\end{equation}
\noindent By combining Eqs.~\eqref{eq:model_evap_flux}, \eqref{eq:model_flux_definition} and~\eqref{eq:model_fraction_derivative}, a differential equation can be obtained for the evolution of water mole fraction in the film:
\begin{multline}
    \frac{1}{(1-x_\text{w})^2} \frac{\mathrm{d} x_\text{w}}{\mathrm{d} t} = - \frac{2\pi D \rho_\text{air} R K}{N M_\text{air}} (a_\text{w} - \mathrm{RH}_\infty) \times \\ \left[a_1\left( \frac{gR^3}{\nu^2}\right)^{1/5} \times \left(\frac{a_\text{w} - \mathrm{RH}_\infty}{\mathrm{RH}_\infty}\right)^{1/5} + a_2\right]. 
\end{multline}
\noindent Introducing a characteristic time $\tilde{\tau} = N M_\text{air} / 2\pi D \rho_\text{air} R K$ and $A_1 = a_1 (gR^3/\nu^2)^{1/5}$, an equation for the evolution of $x_\text{w}$ is obtained:
\begin{equation}
    \frac{1}{(1-x_\text{w})^2} \frac{\mathrm{d} x_\text{w}}{\mathrm{d} t} = - \frac{a_\text{w} - \mathrm{RH}_\infty}{\tilde{\tau}}  \left[A_1 \left(\frac{a_\text{w} - \mathrm{RH}_\infty}{\mathrm{RH}_\infty}\right)^{1/5} + a_2\right]. 
    \label{eq:film_full_model}
\end{equation}
\noindent Provided a model relating activity $a_\text{w}$ of water in the mixture and water fraction $x_\text{w}$ is known, such an equation can be solved numerically.

\section{Activity of water/fructose and water/glycerol mixtures}
\label{sec:app_activity}

Resolution of the full model given by Eq.~\eqref{eq:film_full_model} requires to describe water activity in the ternary water/fructose/glycerol mixture. While both water/glycerol and water/fructose binary mixtures are well studied and described in the literature, the ternary mixture has never been studied.

As both binary mixtures can be well described using the Unifac equation of state~\cite{peres_1996,marcolli_2005}, an attempt was made to use Uniquac predictive modelling, but result failed, probably due to the absence of data on the binary fructose/glycerol mixture.

\begin{figure}[!h]
    \includegraphics[width=0.8\columnwidth]{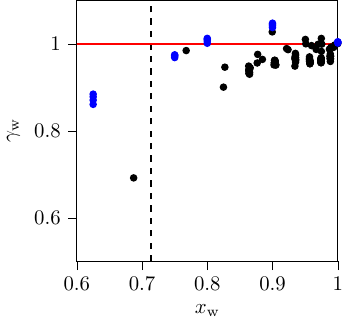}
    \caption{Evolution of activity coefficient $\gamma_\text{w} = a_\text{w}/x_\text{w}$ of water in water/fructose (black symbols) and water/glycerol (blue symbols) with water mole fraction $x_\text{w}$ for the range studied in experiments (typically $x_\text{w} > 0.7$). The black dashed line represents the saturation of water/fructose solution. The red line corresponds to the value $\gamma_\text{w} = 1$ for an ideal mixture.}
    \label{fig:idealite_melange_tot}
\end{figure}

Activity coefficients $\gamma_\text{w} = a_\text{w}/x_\text{w}$ of water in binary water/glycerol and water/fructose mixtures can be obtained from tabulated values of saturation pressure, collected from the Dortmund Data Bank\footnote{Dortmund Data Bank, 2023, www.ddbst.com}. As can be seen on Fig.~\ref{fig:idealite_melange_tot}, in the range of water mole fraction considered in the experiments, both coefficients are rather constant and close to the ideal value $\gamma_\text{w,id} = 1$. Consequently, the hypothesis that the ternary mixture is ideal in the experimental conditions seems acceptable, at least as a first approximation. Also, it is to note that tiny amounts of surfactants (initial mole fraction around $\SI{e-4}{}$) are present in the solution and not considered in the modeling.

\section{Evolution of film viscosity in different conditions}
\label{sec:app_various_initial_conditions}

\begin{figure}[!h]
    \includegraphics[height=5cm]{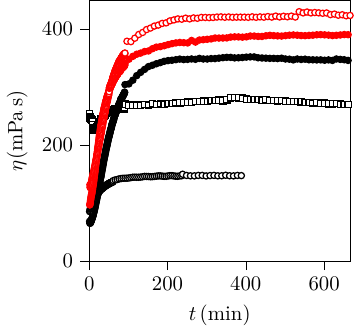}
    \caption{Evolution of averaged viscosity on the observation zone, in the middle of the film with a $20 \times$ objective lens and for different conditions. Different symbols correspond to different initial composition of the film ($\bullet$ : $w^0_\text{f} = 0.53$ and $w^0_\text{g} = 0.06$ identical to main text Fig.~6; $\circ$ : $w^0_\text{f} = 0.60$ and $w^0_\text{g} = 0.05$; {\tiny $\blacksquare$} : $w^0_\text{f} = 0.66$ and $w^0_\text{g} = 0.05$). Different colors correspond to the saturated salt solution used to stabilize humidity away from the film (black : $\mathrm{NaCl}$ ; red : $\mathrm{MgCl_2}$).}
    \label{fig:visco_variousCI_SI_tot}
\end{figure}

Experiments of film evaporation presented in Sections 3.1 and 4.1 of the main text were reproduced with different initial conditions, either by increasing initial fructose fraction of the solution, or by changing the salt solution used to stabilize humidity. Representative examples of viscosity evolution are displayed in Fig.~\ref{fig:visco_variousCI_SI_tot}.

Overall, initial fructose fraction (corresponding to different symbols) has an impact only on the early stage of film evaporation, while final state is mostly imposed by external humidity. Saturated solutions of $\mathrm{MgCl_2}$ (water activity $a_\text{w} = 0.331$ at $\SI{20}{\celsius}$) were expected to reach saturation of fructose solution, and drier final state than solutions of $\mathrm{NaCl}$ (water activity $a_\text{w} = 0.755$ at $\SI{20}{\celsius}$). While final viscosity of the film was indeed higher with $\mathrm{MgCl2}$, the corresponding estimated humidity was not in agreement with the expectations, and no crystallization was observed. This may be due to the limited efficiency of the setup to impose relative humidity (as suggested in the main text), or to improper humidity estimation as it relies on thermodynamic models for water activity in the liquid.

For all experiments, the model detailed in Section~\ref{sec:app_model} gave acceptable results and corresponding fitting parameters are given in Table~\ref{tab:table1}. It is not easy to draw any systematic conclusion from these values, aside from consistent orders of magnitude.

\begin{table}[!htb]
\centering
\begin{equation}
\begin{array}{|c|c|c|c|c|c|}
\hline
w^0_\text{f} & w^0_\text{g} & \text{Salt} & \mathrm{RH_{\infty}} & \tilde{\tau} \, (\SI{}{\minute}) & e_\text{film} \, (\SI{}{\micro\meter}) \\ 
\hline
\multirow{2}{*}{0.53} & \multirow{2}{*}{0.06} & \mathrm{NaCl}   & 0.731 & 3.8 & 400 \\ \cline{3-6}
                      &                       & \mathrm{MgCl_2} & 0.722 & 3.4 & 360 \\ 
\hline
\multirow{2}{*}{0.60} & \multirow{2}{*}{0.05} & \mathrm{NaCl}   & 0.816 & 1.4 & 130 \\ \cline{3-6}
                      &                       & \mathrm{MgCl_2} & 0.748 & 3.8 & 370 \\ 
\hline
0.66                  & 0.05                  & \mathrm{NaCl}   & 0.788 & 3.3 & 350 \\
\hline
\end{array} 
\end{equation}
\caption{Estimated final humidity $\mathrm{RH_\infty}$, characteristic time $\tilde{\tau}$ of the model and estimated initial film thickness $e_\text{film}$ for different initial conditions (initial fructose and glycerol mass fractions $w^0_\text{f}$ and $w^0_\text{g}$, and type of saturated salt solution).} 
\label{tab:table1}
\end{table}

\end{document}